# IRIS: A Low Duty Cycle Cross-Layer Protocol for Long-Range Wireless Sensor Networks with Low Power Budget

Yi Chu, Paul Mitchell, David Grace, Jonathan Roberts, Dominic White and Tautvydas Mickus

*Abstract*—**This paper presents a cross-layer protocol (IRIS) designed for long-range pipeline Wireless Sensor Networks with extremely low power budget, typically seen in a range of monitoring applications. IRIS uses ping packets initiated by a base station to travel through the multi-hop network and carry monitoring information. The protocol is able to operate with less than 1% duty cycle, thereby conforming to ISM band spectrum regulations in the 868MHz band. The duty cycle can be flexibly configured to meet other regulations/power budgets as well as to improve the route forming performance. Simulation results show guaranteed route formation in different network topologies with various protocol configurations. System robustness against unreliable wireless connections and node failures are also demonstrated by simulations.**

## I. INTRODUCTION

Wireless Sensor Networks (WSNs) offer low-cost solutions for long-term monitoring tasks operating in various types of environments [1]. Compared with conventional wired monitoring systems, wireless sensor nodes can be rapidly deployed with minimal infrastructure requirements, with the potential to automatically form a network according to their self-organising nature. During operation, WSNs can be designed to be robust to single node failures thereby being exempt from frequent maintenance by human engineers. In recent decades, WSNs have been applied to many applications, such as geological event monitoring [2], animal habitat monitoring [3], resource industry [5], health monitoring [6], smart cities [7], smart grids [8] and smart farming [9]. Now WSNs have become the solid foundation of the rapidly expanding Internet of Things (IoT) [10].

Wireless sensor nodes are normally powered by batteries or energy harvesting devices [11], which makes energy efficiency a critical requirement where WSNs need to operate for a long period of time. The Radio Frequency (RF) module on a node is usually a major energy consumer, and the power consumption is mainly from (re)transmitting, receiving packets and idle listening. Energy consumed by packet collisions, retransmissions, exchanging control information, idle listening and overhearing are considered as overheads [12]. Well-designed Medium Access Control (MAC) and network layers should be able to keep these overheads as low as possible, while achieving the performance required by the application. For example S-MAC [13] and Z-MAC [14] use duty cycling procedures to switch nodes between active and sleep states to conserve energy. Some regulations have also considered limiting the duty cycles for spectrum sharing purposes. For example the UK regulator Ofcom's IR 2030 document [15] has limited the duty cycle of most 868 MHz industrial, scientific and medical (ISM) bands to 1%, where such a band is widely used by Long-Range Wide-Area Network (LoRaWAN) nodes for many IoT applications [16].

Having a low duty cycle benefits the energy efficiency while meeting the regulations, however it limits the number of packet exchanges between nodes, which are necessary for operating the MAC and network layer protocols. Motivated by these constraints, in this paper we present a simple but novel cross-layer protocol (IRIS) which is designed to operate with an ultra-low duty cycle, for long-range monitoring tasks such as river/canal monitoring, coastline monitoring, underwater cable, motorway and railway monitoring. These applications require pipeline network topologies (potentially over hundreds of kilometres), often in environments that are not well served by existing wireless infrastructure, so the monitoring information must be passed over a large number of hops to reach the destination. The organisation of routes through which information travels becomes challenging in such situations, where lengths of the pipelines are long and involve a large number of nodes. The proposed IRIS protocol integrates energy efficient MAC and network layers, achieved by simple node logic to achieve operation with a less than 1% duty cycle. Specific contributions of this paper are summarised as follows:

1) A protocol which can operate in a scenario where the duty cycles of all nodes is less than 1% during both network initialisation and normal operation. Unlike many state-of-the-art protocols which require the nodes to have high duty cycles during initialisation, IRIS can be deployed to nodes with extremely limited initial energy storage (e.g. nodes powered by energy harvesting devices only without batteries).

2) A protocol where the duty cycle can be controlled. IRIS uses ping packets periodically initiated by a base station at one end of the network (and relayed by multiple nodes on a route) to carry monitoring information to a base station at the other end of the network. This allows the base station to control the duty cycle of the network according to different requirements. Ping packets are widely used in networks using the Internet Protocol (IP) to test the reachability to certain destinations. In the IRIS protocol the ping packets will ultimately reach the end base station of the WSN.

3) An energy efficient route discovery process. Route finding is completed while relaying the ping packets without any knowledge of neighbour nodes. The simple nature of IRIS allows the protocol to be scaled to networks with different numbers of nodes and hops. The low computational requirements make the protocol feasible for deployment on low-cost nodes.

4) A protocol which integrates coordinated MAC layer and network layer functions under energy and bandwidth constraints. The node logic is able to achieve energy efficient transmission/reception scheduling while relaying the monitoring information towards the base station and to adapt to network topology changes. The simple nature of the protocol allows the network to operate with extremely low date rate available.

The rest of the paper is organised as follows. Section II presents the related work on low duty cycle protocols. Section III introduces the network topology and the main idea of IRIS. Section IV explains the methods IRIS uses to achieve reliability. Section V presents the simulation results and discussions. Section VI concludes the paper.

## II. RELATED WORK

Duty cycling is widely used by many MAC protocols to switch wireless sensor nodes into a lower power sleep state whenever possible to conserve energy. For example, S-MAC [13] sets up synchronised sleep schedules across a neighbourhood to make sure they all wake up at the same time to exchange packets. During active periods, nodes use Carrier-Sense Multiple Access with Collision Avoidance (CSMA/CA) with Request to Send (RTS)/Confirm to Send (CTS) handshakes to contend for transmissions. Many variants [17] of S-MAC have been developed to solve the restriction of a fixed duty cycle by considering different service requirements to improve the energy efficiency, but reaching ultra-low (e.g. 1% or lower) duty cycle operation remains difficult.

Another MAC protocol with synchronised duty cycle Scheduled Channel Polling (SCP) [22] has achieved an operating duty cycle of less than 1% in optimal conditions by significantly reducing the length of preambles used in Low-Power Listening (LPL). However SCP-MAC requires considerable energy during initialisation for the purpose of synchronisation. X-MAC [23] has tried to tackle the same problem of LPL by using multiple short preambles and acknowledgement (ACK) from the receiving node to initiate transmissions earlier without synchronised duty cycles. Duty cycles of 5% to 10% are achieved by X-MAC and higher overheads are incurred to compensate for the asynchrony. Another asynchronous protocol AP-MAC [24] uses randomly transmitted beacons to broadcast node sleep schedules to neighbours, so that the sender can predict the receiver's next active period to reduce overheads. However AP-MAC also requires significant initial energy and its lowest duty cycle is approximately 10%.

MAC and network layers working collaboratively benefits the overall performance of a WSN. Many cross-layer protocols are developed to exploit the information from both layers to achieve better performance. For example, RMAC [25] exploits the routing information between source and destination nodes to effectively switch the relay nodes between active and sleep states to save energy. RMAC also utilises the broadcast nature of wireless signals by sending a single Pioneer Control Frame (PION) message to replace the ACK to the previous hop and the RTS to the next hop. RMAC has achieved a 2% to 3% duty cycle during normal operation but the route finding is not included. Light-Weight Opportunistic Forwarding (LWOF) [26] uses the location information of the neighbour nodes to efficiently forward packets to the sink. However the duty cycle is relatively high (5% to 18%) because of the preambles of LPL and Global Positioning System (GPS) modules, both of which are not always available for WSNs. Dynamic Switching-Based Reliable Flooding (DSRF) [27] has developed the Automatic-Repeat-Request (ARQ) based flooding tree by using the topology information of parent, sibling and child nodes to align their sleep schedules to reduce overhearing, thereby improving the energy efficiency. DSRF has achieved a 1% duty cycle with the assumption of knowing the duty cycles of all neighbour nodes. The authors of [27] have later proposed a distributed Minimum-Delay Energy-efficient flooding Tree (MDET) [28] network layer algorithm to construct an energy optimal flooding tree. However energy efficiency is not considered during route finding and the nodes need to be active when constructing the flooding tree.

Although some state-of-the-art protocols are able to achieve low duty cycle while operating in steady state, the consistently low energy cost during all phases of operations is yet to be addressed. High energy consumption during network initialisation and route finding will significantly reduce the applicability of the protocols to networks without sufficient amount of initial energy storage (e.g. networks powered by energy harvesting devices). The IRIS protocol tackles the problem by maintaining the ability to operate under low energy budget across the lifespan of the network. This unique feature initiates the possibility of further reducing the cost of the WSNs (batteries are not required) while maintaining long life time.

## III. IRIS PROTOCOL DESIGN

Long-term and long-range monitoring tasks are always challenging for WSNs because of the constraints of energy, processing power and connectivity. Motivated by these constraints, IRIS uses a simple but novel approach to complete these tasks with extremely low energy availability. This section describes the network topology of the applications that IRIS is designed for, and how IRIS integrates the MAC and network layers.

### A. Network Topology

The target applications of IRIS have a pipeline type of topology such as rivers, coastlines, underwater cables, motorways and railways. For the purpose of generality, the low-cost nodes are assumed to be randomly deployed along the pipeline and the monitoring information from all source nodes is forwarded to a base station (or a control centre) at the end of the network over a large number of hops. A second base station is deployed at the other end of the pipeline for the purpose of initiating communication from nodes along the pipeline. The base stations are located far away from each other, but they have direct links to Internet gateways so that the monitoring information can be sent to a remote control centre. In practice, with the exception of a more substantial power source, the base stations could employ the same technology as the low cost nodes along the pipeline, because the required processing capability of the base stations is not significantly different from other nodes in the network. Infrastructure (e.g. power, network access) is only required at the base stations and the rest of the network is able to operate independently. The IRIS protocol is designed to allow the network to operate without any prior neighbourhood information and without any pre-determined structure. The low cost nodes have no knowledge of their geolocations. The nodes can be deployed in any physical order at random locations. The only fundamental requirement is the need for nodes to be deployed within reasonable radio range such that end-to-end connectivity is achievable. Fig. 1 shows an example of the proposed WSN.

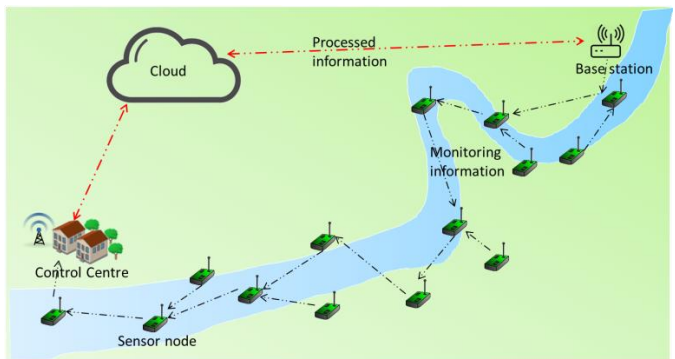

Fig. 1 Network topology

### B. MAC Layer Design of the IRIS Protocol

IRIS uses ping packets initiated by the base station to propagate through the pipeline network while collecting monitoring information from the nodes who can hear the pings. Time Division Multiple Access (TDMA) is used to make sure the nodes only switch to an active state when necessary to conserve energy. We refer to the time period of one duty cycle as a frame and the frame length is the same across all nodes in the network. One frame is further divided into a number of slots, where the length of a slot allows a node to send a ping packet and receive an ACK packet. The slot length varies according to the different requirements, and the frame size can be adjusted based on the power availability and the duty cycle requirements.

Fig. 2 shows an example illustration of the MAC layer of the IRIS protocol. This example shows a 6-node section in the middle of a multi-hop WSN. Nodes A, D and F which relay the ping packets are defined as route nodes, and nodes B, C and E which report monitoring information to the route nodes are defined as non-route nodes. Before the example starts we assume that node A has already received a ping packet from its previous hop route node. In slot 1 node A sends a ping packet to node D and receives an ACK packet. Node A keeps listening in slot 2 in case other nodes want to report their monitoring information after hearing the ping packet (node B could report in this example). Node A switches to a low power sleep state after the listening period. Node D can switch to the sleep state in slot 2 however this may be inefficient for some hardware designs because node D will be active again in slot 3. In slot 3, node D sends the ping packet to node F and receives an ACK. Node E hears the ping packet and has some information to report, so it sends a report packet to node D in slot 4 and receives an ACK. Node D switches to the sleep state after slot 4. A route node wakes up in the same slot every frame and expects a ping packet from the previous hop. In practice a route node can wake up slightly earlier and adjust its local timer according to the received ping packet to compensate the potential clock drift.

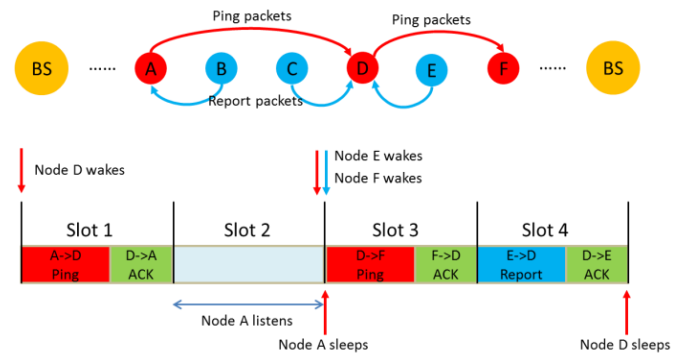

Fig. 2 IRIS MAC layer example

The node which has not heard a ping packet is defined as a searching node. A node normally starts as a searching node after powering on or losing connection to an existing network. For a searching node to join an existing network, it starts by randomly selecting a slot to listen to. The number of slots it continues listening in depends on the energy available. If it does not hear any ping packets during the active period in the current frame, it shifts its active period forward in the next frame. The same process applies when the first node joins the base station, and the base station that initiates the ping packets can be considered as the first route node. If a node listens to slots 1 to 4 in the first frame and does not hear any ping packets, it listens to slots 5 to 8 in the next frame. Once it hears a ping packet, it can either latch on to the sender of the ping in every frame or keep shifting the active period forward until it cannot hear a ping packet then loop the active period within the slots that it can hear ping packets. For example in Fig. 3 the searching node is active in slots 9-12 in frame 1 and hears a ping packet in slot 11 then becomes a non-route node. In frame 2 it can either stay active in the same slots 9-12 (option 1) or shift the active period to slots 13-16 (option 2). In the former case the node only reports to one route node and in the latter case the node can potentially report to multiple route nodes which could benefit the contention.

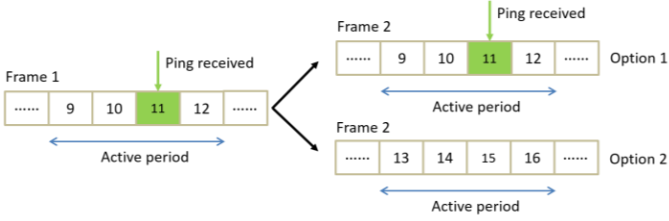

Fig. 3 Example of a non-route shifting its active period

## C. Network Layer Design of the IRIS Protocol

The only information required by IRIS to find a route between two base stations is a unique node ID configured before deployment and there are no specific requirements during the deployment (i.e. it is not necessary for the nodes to be deployed in any particular order and location information is not required). The only information the nodes require is the pre-programmed slot duration, the number of slots per frame and the number of slots they should remain active for, all which stem from the system design for a particular deployment to meet regularity and energy budget requirements. Energy consuming neighbour discovery is not needed by IRIS. The route formation starts with one base station at one end of the network broadcasting a ping packet. The timing of this ping packet transmission defines the timing (and start) of a frame. Fig. 4 shows the structure of the ping packet, which includes the packet type, ID of the sender and destination (next-hop) nodes, a binary link ID which indicates whether the route is formed, and the payload reserved for monitoring information. If the sender does not have an intended destination node, it just includes "-1" in the field of destination ID indicating that it is looking for a next hop to join the network.

| Pk type | Sender ID | Dest ID | Link ID | Payload |
|---------|-----------|---------|---------|---------|

Fig. 4 Ping packet structure

While the base station is broadcasting ping packets in the first slot of every frame, other nodes wakeup at random times and listen for the defined active period (a specific number of slot durations) and then go back to sleep for the defined sleep period. If a node does not hear anything during its active period, it delays its active period such that it wakes up a certain number of slots later in the next frame. This process allows all nodes to search for a travelling ping packet without any knowledge about the network topology or timing. Once a node hears a ping packet with "-1" destination ID (the sender is defined as a route-end node), it replies with an ACK (including its own ID) to declare its interest in joining the route to relay ping packets. In the next frame, it configures its active period to start in tandem with the timing of the ping packet it received and expects another ping. If the destination ID is still "-1", it configures its active period to start with a random slot in the next frame to avoid potential ACK collisions. If the destination ID is not "-1" nor its own ID, it becomes a non-route node and follows the activities as described in Fig. 3. If the destination ID of the ping packet now becomes its own ID, it means the sender has selected it as the next hop and it becomes the new route-end node. Then the new route-end node repeats the behaviour of its previous hop node by sending ping packets with a "-1" destination ID and expecting other nodes to join the route. This process repeats hop-by-hop until the base station at the other end of the network receives a ping. Then both base stations know that a route has been formed and the next ping packet initiated by the base station will have "1" in the link ID field indicating that the route has been formed. This also indicates that the non-route nodes (e.g. nodes B, C and E in Fig. 2) can start to report their monitoring information periodically to the route node after hearing a ping. The basic operating states of the base station which originates the ping packets is shown in Fig. 5. The basic operating states of all other nodes are shown in Fig. 6. Note that only the base station can change the Link ID field in the ping packets.

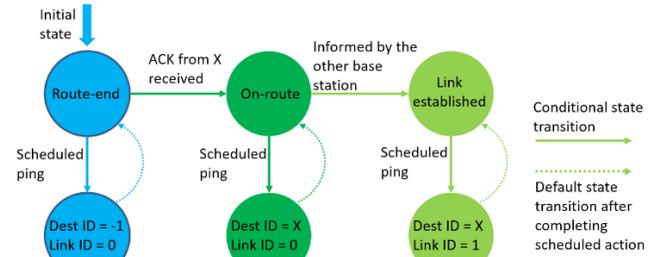

Fig. 5 Base station operating states

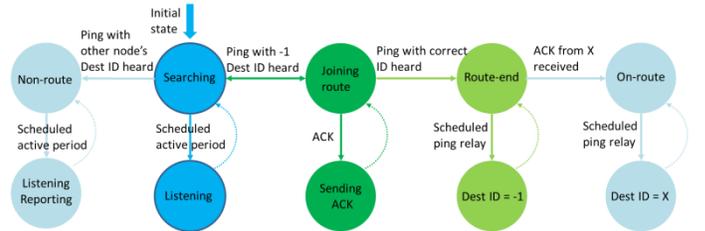

Fig. 6 Node operating states

## IV. IRIS PROTOCOL RELIABILITY METHODS

The node logic described in the previous section establishes the basic MAC and network layer operations. However additional features are needed to secure the reliability requirements. This section demonstrates the methods IRIS uses to avoid routing loops, to make sure the route propagates forward and to achieve robustness against unreliable wireless connections.

### A. Inefficient Routing Avoidance

First of all, one node is able to relay at most one ping packet per frame given its limited active period (forced by the low duty cycle operation), so creating a loop via the same node is naturally impossible for IRIS. However, without information about neighbour nodes and their locations, it is possible for the route to geographically loop or even propagate backwards towards the base station which initiates the ping packets. For example, in Fig. 7 the route geographically propagates backwards at node D and E.

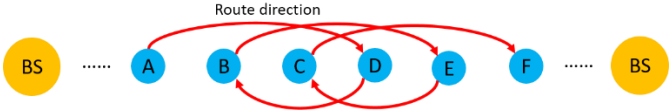

Fig. 7 Some hops propagate backwards

To avoid such inefficient routes, the density of the route nodes can be controlled by preventing a node from joining the route when it hears a lot of activity in the neighbourhood. When a node in a searching state shifts its active period forward in each frame, it counts the number of ping packets with non "-1" destination ID it has heard. If this number exceeds a certain threshold, it will not attempt to join the network. We define this threshold as connection limit (*conlimit*), and it is configurable. For example, in Fig. 8, node B is able to hear the ping packets addressed to nodes C and D, and also the ping packet with −1 destination ID from node D. If the *conlimit* is set to 2, node B will not attempt to join the route, thereby avoiding the potential of creating an inefficient route. This *conlimit* also helps finding the next hop relatively far away geographically, thereby reducing the total number of hops needed to reach the other base station.

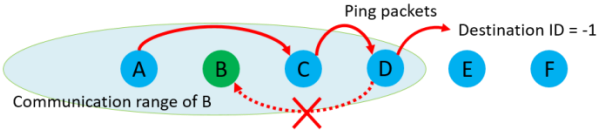

Fig. 8 Example of *conlimit* = 2

In the situation where a route-end node is not able to find a next hop to propagate the ping packets (caused by either some neighbour nodes being limited by *conlimit* or because a node does not have a next hop node within radio range), the node sends a drop packet to its previous hop and switches to the searching state. Then the previous hop node becomes the route-end node which broadcasts pings "−1" destination ID to try and find an alternative next hop. We set a threshold *frameout* for the nodes to decide when to send the drop packet. The threshold *frameout* is defined by the number of consecutives frames that the route-end node cannot find a next hop node. For example in Fig. 9 when node C is the route-end node, nodes D and E can hear two ping packets with non-"-1" destination when they are active thereby will not attempt to become route nodes. When *frameout* of node C exceeds a certain value it sends a drop packet to node B to find an alternative route.

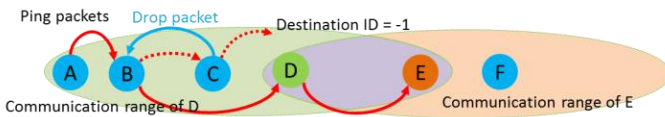

Fig. 9 Example of the drop packet

### B. Robustness against Unreliable Wireless Connections

In a practical environment, the wireless connections can be subject to interference from other devices using the same spectrum, or the channel could be in a deep fade. Either condition will cause the packets to be partially received/damaged or completely lost. Moreover the low-cost nodes can stop functioning due to a flat battery or physical damage, which can also cause packet losses. The actions that IRIS will take according to the losses of different types of packets are as follows:

1) **Ping packet loss**: a node does not hear a correct ping (with destination ID) when it expects one. If the node is a route-end node, missing a ping packet could indicate either the previous hop is not functioning, or it is simply experiencing a bad connection. The route-end node will generate a ping packet and send it to the next hop as if it heard the ping packet from the previous hop, so that the downstream route is still kept alive in case of minor ping packet losses. At the same time it increases a counter called *phq_frameout* (previous hop quiet) which indicates the number of consecutive frames that the route node misses ping packets. Once the counter exceeds a certain threshold, the route (-end) node stops generating ping packets and switches back to the searching state. If the node is a non-route node which has latched onto a route node to send its report packets, it will increase a counter called *rq_frameout* (route quiet) in a similar fashion and switch to searching state once the counter expires.

2) **ACK (which replies to a ping with destination ID) packet loss**: a node does not hear an ACK from the next hop after it sends a ping. This indicates that either the next hop is not functioning or there is a bad connection. The route node increases a counter called *nhq_frameout* (next hop quiet) and switches to the route-end state once the counter expires and starts sending ping packets with "-1" destination ID.

3) **ACK (which replies to a ping with "-1" destination ID) packet loss or collision**: an ACK is lost or collided when a searching node attempts to join the route. This does not affect the route-end node which originates the ping packet. If the searching node which sends the ACK wakes up in the same slot next frame and hears the ping with "-1" destination ID again, it configures its active period to start with a random slot in the next frame.

4) **Report packet and its associated ACK losses**: a non-route node may experience collision or channel fading when sending the report packet to the route node, or the route node sending the ACK back. These do not affect the sleep schedule of the non-route node. Its active period remains as long as it can hear the ping packets.

5) **Drop packet and its associated ACK losses**. The route-end node will keep sending the drop packets in the same slot every frame until it is successfully acknowledged if it can hear the ping from the previous hop.

With all the aforementioned features, the node operating states shown in Fig. 6 are updated, as shown in Fig. 10. The node (especially route nodes) failures can be considered as "permanent" packet losses. The failure of a route node will cause the downstream route to be reconstructed by following the operating states in Fig. 10 while the upstream route remains functioning.

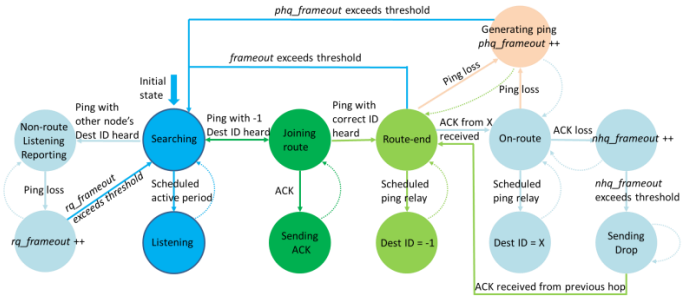

Fig. 10 Updated node operating states

## V. SIMULATIONS AND DISCUSSIONS

In this section we evaluate the performance of IRIS with Monte Carlo simulations using Matlab. The network layer metrics to be evaluated include the time required to establish a route and to recover from node failures, and the robustness against clock drift. The MAC layer metrics to be evaluated include the throughput and delay for the monitoring information to reach the base station after the route is established, and the robustness to unreliable wireless connections.

### A. Simulation Parameters

We assume the nodes are equipped with the LoRa transceiver Semtech SX1276 [29] and the Texas Instruments MSP430F2617 microcontroller [30]. The LoRa transceiver operates in the suggested low-power, low-bit-rate and long-range mode. The properties of the LoRa transceiver are listed in TABLE I. The microcontroller has a current draw of 5.84 mA when active at 16 MHz (could be lower for slower processor speed) and 0.5 µA in standby mode. The nodes are also equipped with a super capacitor charged by an energy harvester to power the transceiver and the microcontroller. We consider that a centimetre-level size wind energy harvester is used such as the one presented in [31] which is able to produce 1.7 mA current draw at a wind speed of 5 m/s. There are also many other similar options available as presented in [32]. These can be tailored to a specific environment to provide a sufficiently reliable power source. A nominal 5 F super capacitor is used and it can be charged to 4.1 V. An output regulator (with 60% efficiency) cuts off when the capacitor is discharged to 2.25 V, so there is 1.02 mA current and 3 F capacitor storage available to power the node. The node actions and the associated charge gained/consumed are summarised in TABLE II.

For a node can transmit for 1 second, it needs to be charged for approximately 49 seconds while asleep (equivalent to a 2% duty cycle). For a node to receive/listen for 1 second, it needs to be charged for approximately 16 seconds (equivalent to a 6% duty cycle). The energy budget indicates that the nodes can easily operate at 1% duty cycle (or below) to meet the regulations.

The ping packet uses a short packet size with 22-byte payload, 4-byte preamble and 2-byte Cyclic Redundancy Check (CRC). The LoRa calculator [29] shows that 297 ms is required to transmit the ping packet and approximately 200 ms to transmit the ACK. It costs a node 14.8 mC to send a ping and

3.33 mC to receive the ACK in one slot, and its next hop node needs 4.49 mC to receive the ping and 9.97 mC to send the ACK. According to the time needed to send a ping and an ACK, we define the slot length as 500 ms and it costs 7.82 mC to listen during a slot. The most power consuming nodes in the network are the route nodes, which need to be active for at least 3 slots (ping reception, ping transmission, listening/report reception) during one duty cycle. For example in Fig. 2 the 3 active slots of node D require 14.46 mC, 18.13 mC and 14.46 mC of charge respectively (report packet has the same length as ping). To compensate the energy consumption of during the active period of node D approximately 46 s of charge time is needed, which makes a maximum affordable 3.3% duty cycle. If we assume node D is active in slot 2 (listening mode) as well, the maximum affordable duty cycle becomes about 3.7% (roughly 54 s of charge to support 4 active slots). As described earlier, we consider each route node to have 1 slot

### TABLE I
### LoRa TRANSCEIVER PROPERTIES

| Property | Value |
|---|---|
| Spreading factor | 10 |
| Bandwidth | 125 kHz |
| Coding rate | 4/5 |
| Data rate | 976 bps |
| Transmitting power | 14 dBm |
| Transmitting current draw | 44 mA |
| Receiving current draw | 10.8 mA |
| Sleep mode current draw | 0.2 µA |
| Frequency | 868.1 MHz |

### TABLE II
### CHARGES AND NODE ACTIONS

| Action | Charge/s |
|---|---|
| Charging | +1.02 mC |
| Sleeping | -0.7 µC |
| Transmitting | -49.84 mC |
| Receiving/listening | -16.64 mC |
| Processing | -5.84 mC |

reserved for non-route nodes to send report packets. This number could be further extended given an improved energy budget. However, the bottleneck is the limited storage space in ping packets, so having more frequent reports could be unnecessary. With this configuration, we define all nodes to have a 4 slot active period per frame, and each frame has 400 slots (200 seconds) which meet the 1% target duty cycle (active). So it takes 100 frames (about 5 and half hours) for a searching node to explore every slot. Table III summarises the duty cycles of the nodes with different tasks in the network. The maximum affordable active column shows the upper bound of

### TABLE III
### NODES AND DUTY CYCLES

| Node type | Duty cycle (active) | Transmission only | Maximum affordable active |
|---|---|---|---|
| On-route (sending ping and receiving report) | 1% | 0.35% | 3.7% |
| On-route (sending ping) | 1% | 0.25% | 4% |
| Non-route (sending report) | 1% | 0.15% | 4.7% |
| Non-route (searching) | 1% | 0% | 6% |

the duty cycle while exploiting all the energy budget.

### B. Route Formation of Different Networks

We first evaluate the route formation capability of IRIS by simulating a hypothetical pipeline network. The network has 300 nodes including two base stations at both ends. The nodes are deployed 500 meters from each neighbour and the total network length is 150 km. We assume that the LoRa devices are set to long-range mode in which they have a 20 km communication range with a line-of-sight (LoS) path (the record is 702 km [33]). The first sets of simulations are based on this configuration in order to find an appropriate combination of the IRIS protocol parameters mentioned in section IV. A (*frameout* and *conlimit*). Both parameters have significant impact on IRIS performance. Setting a large *frameout* may cause a route-end node to return to the searching state very late when it has no other nodes in range to become a next hop. However if *frameout* is too small, the route-end node may miss hitting the active period of a potential next hop node in the search state, given the ultra-low duty cycle. Having a large *conlimit* may create unnecessary hops in the route and increase the end-to-end delay of ping packets. On the other hand, a small *conlimit* decreases the probability of finding a next hop before *frameout* expires which increases the time needed for route formation. Monte Carlo simulations are necessary to discover the impact of different parameter configurations and the sensitivity of the system to the settings.

Fig. 11 to Fig. 13 show the Cumulative Distribution Function (CDF) of the time to form a route with different combinations of parameters. The time to form a route is defined as the time from the instant that the first ping initiated by the base station until the time the other base station receives a ping. Each curve represents the results of 50 simulations. This applies to all simulation results that appear later if not specified otherwise. Fig. 11 shows that the route formation is not sensitive to *frameout* when *conlimit* is set to 1. Fig. 12 indicates that *frameout* = 50 performs other values when *conlimit* is set to 2. Fig. 13 shows that a *conlimit* = 1 is slightly better than larger *conlimit* values. The first impression is that the combination of *frameout* = 50 and *conlimit* = 1 is better than other settings. Generally in 90% (or more) of the simulations of any parameter combination, a route can be found within less than 7 hours and the number of hops are within the range from 11 to 13 hops. Once a route is found, the end base station will receive one ping packet which traverses through all route nodes with 22-byte payload every frame. This also indicates that the network throughput is 396 bytes/hour. Note that the number of hops does not affect the network throughput as long as the number of slots per frame and the number of active slots per frame are fixed. The network throughput is obtained based on the parameters in Table I and it can be increased by adjusting the spreading factor and bandwidth.

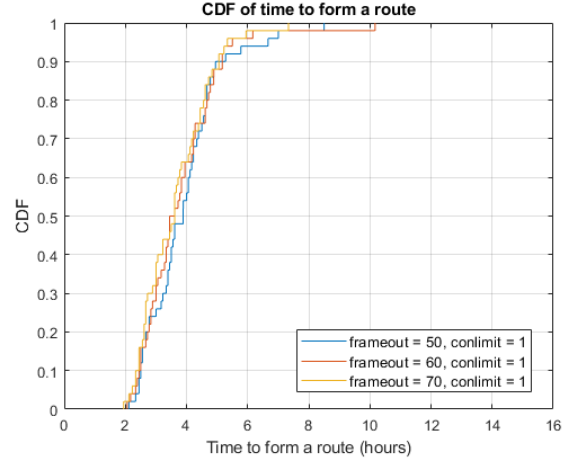

Fig. 11 Results of *conlimit* = 1

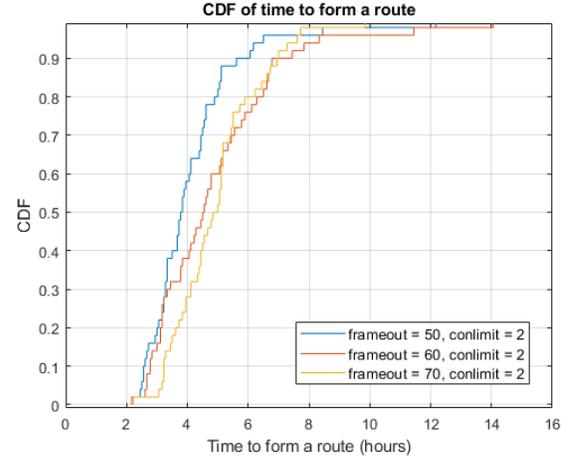

Fig. 12 Results of *conlimit* = 2

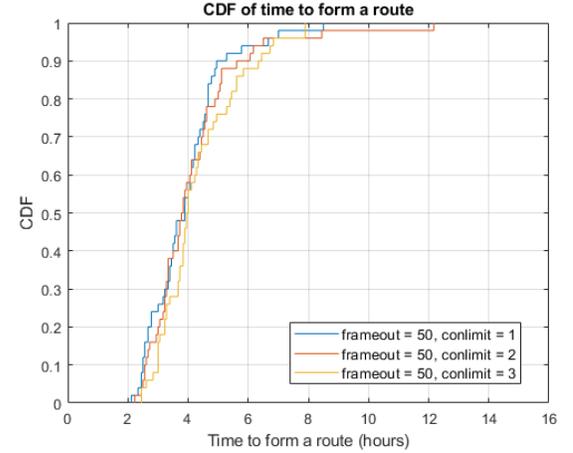

Fig. 13 Results of *frameout* = 50

To validate the performance stability of IRIS with different network topologies, we construct a random pipeline network with 300 nodes. The network is generated node-by-node, with 80% of the neighbour node distances under 2 km and 20% of the neighbour node distances between 2 km and 5 km. The distance between two neighbour nodes follows the uniform distribution. The total length of this random network is approximately 467 km. Fig. 14 shows that the route formation is faster when *conlimit* is set to 1, which is consistent with earlier simulation results. Fig. 15 shows that setting *frameout* =

50 slightly outperforms other *frameout* values, which is also consistent with the first impressions we have earlier. With the best parameter configuration, a route is found within 35 hours for 90% of the simulations, and 50 hours for all simulations. Given the time scale of typical WSN deployments in these long range monitoring scenarios, a 2-day route formation phase is negligible to a WSN which will operate for months or years. The average number of hops of all found routes is about 35, which is much larger than the previous hypothetical network (because of the longer distance between neighbour nodes).

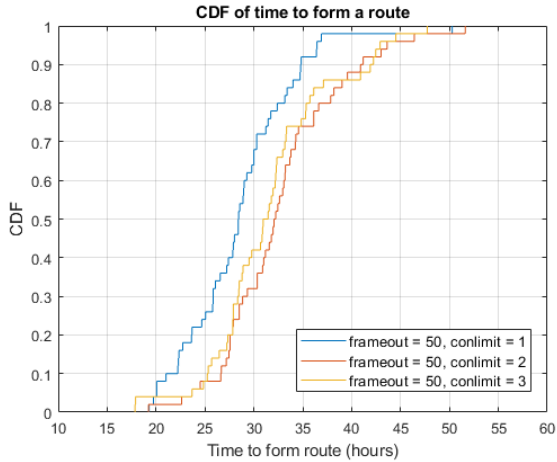

Fig. 14 Results of *frameout* = 50 (random network 1)

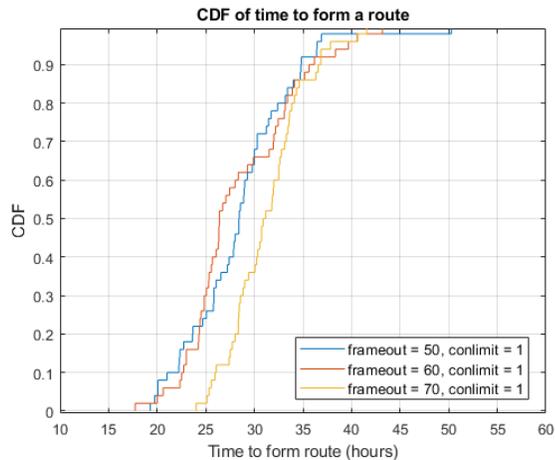

Fig. 15 Results of *conlimit* = 1 (random network 1)

We also construct two other random networks following the same rules to make sure the route finding process operate correctly with different network topologies configurations. Fig. 16 compares the route formation time of 3 networks with 467 km, 423 km and 430 km length respectively. The average numbers of hops are 35, 31 and 32 for all 3 networks. As expected, IRIS can form routes faster for networks with fewer hops and shorter length. The route formation time for 90% of the simulations are 35, 28 and 31 hours for 3 networks respectively. It is clear that IRIS is able to provide reliable route formation for different pipeline networks without any prior topology information while consistently maintaining low energy cost, which is the major advantage of the protocol.

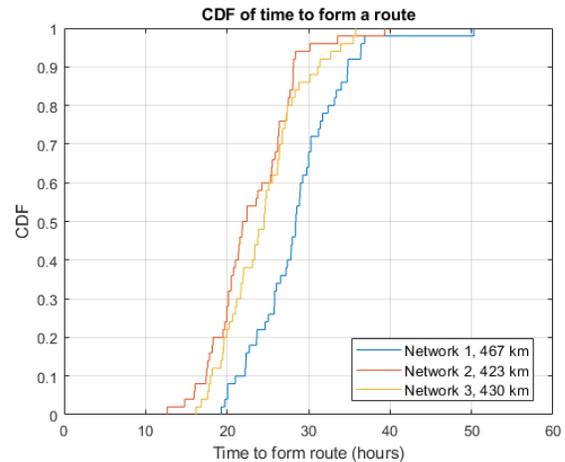

Fig. 16 Results of 3 random networks

## C. Route Formation with Different Wireless Coverage

When a LoRa transceiver fails to reach the 20 km communication range this could be due to many different reasons, IRIS can adapt to the reduced range and find routes with a greater number of hops. Fig. 17 shows the route formation time of the hypothetical network (500 m neighbour distance) with different communication ranges. We can see that the 90the percentile point is increased to 8 and 14 hours with the reduced 15km and 10 km range respectively. Fig. 18 shows a boxplot of numbers of hops of all found routes, where the upper and lower bounds of box represent 75% and 25% of the samples with the red lines as the median value $50^{th}$ percentile (the 10 km plot has no lower bound whisker because it overlaps with the box). The two whiskers of each box are the upper and lower 1.5 interquartile ranges. When the communication range is reduced by half (10 km), the average number of hops is doubled. The route formation time is almost three times as the 20 km range, because number of potential next hop nodes are reduced and the route-end node has to wait longer for them to tune their active periods to receive the ping.

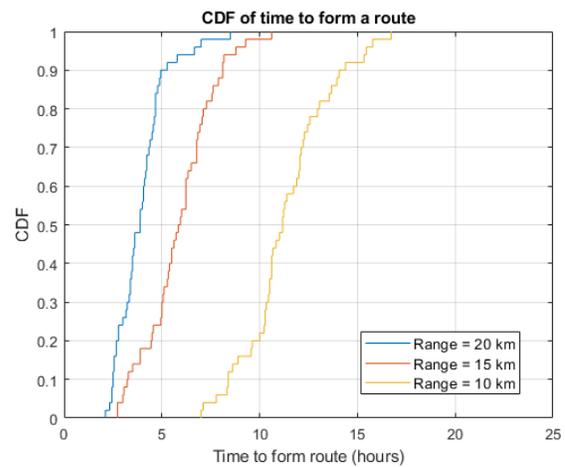

Fig. 17 Hypothetical network with different communication range

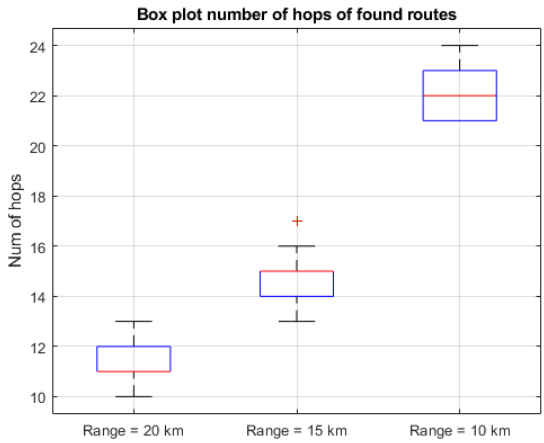

Fig. 18 Hypothetical network with different communication range

Fig. 19 and Fig. 20 show the same results of the first random network (random network 1 in Fig. 14 and Fig. 15) constructed. Similar increasing trends of the route formation time and number of hops can be observed with the reduced communication range. When the communication range is halved to 10 km, it takes 4 times longer to find a route because of much less potential next hop nodes in range. The number of hops to reach the end base station is also almost doubled. These results show the consistently guaranteed route formation with different communication ranges.

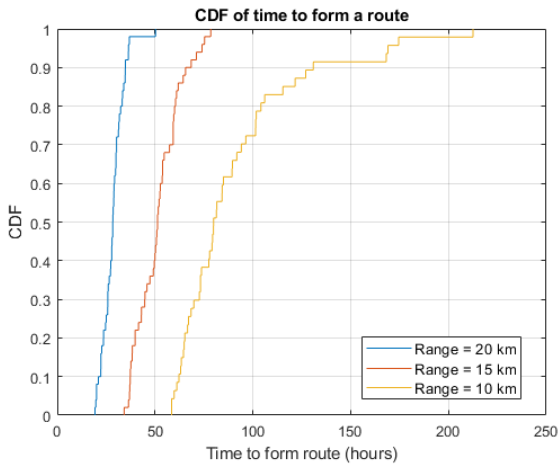

Fig. 19 Random network 1 with different communication range

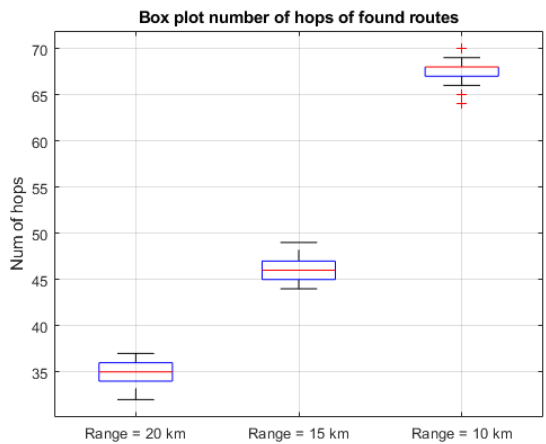

Fig. 20 Random network 1 with different communication range

### D. Robustness against Unreliable Connections and Devices

In previous simulations we assume that all nodes have perfect wireless connections and all packets sent during the active periods of the destination nodes are received successfully. To evaluate how the unreliable connections and packet losses affect the IRIS performance, we simulate the route formation process with different packet loss rates. The times between two successive packet loss events follow either uniform or exponential distributions (Poisson process). From Fig. 21 we can see that the simulations with the same packet loss rate of both distributions have similar performance. 1% packet losses have a negligible effect on the route formation time. 90% of the simulations can find routes within 60 hours even with 10% packet losses, compared with the 35 hours of 0% packet loss. Note that the packet loss rate is applied to all packet types and some actions require two packets to complete (e.g. joining the route).

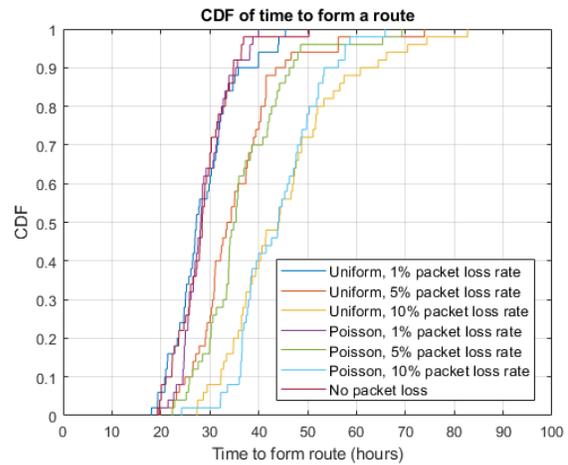

Fig. 21 Route formation time with packet losses

IRIS requires loose time synchronisation to align the active periods of nodes with the associated ping packets. Packet collision is rare due to the unique nature of the actions. Clock drift is a common effect of low-cost oscillators which will cause the timing of slots to slowly shift across different nodes. LoRa devices could have 20 ppm up to maximum 200 ppm clock drift [29], which are 1.73 and 17.3 seconds/day or 4 and 40 ms/frame. The slot length is 500 ms so if a node synchronises its clock to the ping packets every frame and starts its active period slightly earlier with a 50 ms guard period (10% slot duration), the effect of clock drift should be compensated. In Fig. 22 we can see that clock drift only slightly affects the route formation time.

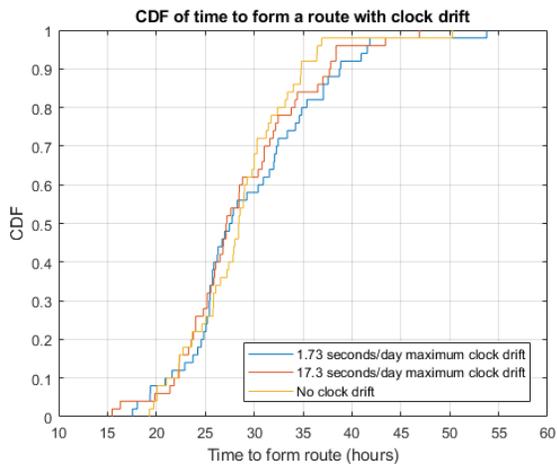

Fig. 22 Route formation time with clock drift

### E. Route Recovery from Node Failure

Node failure is common during the operation of WSNs given the nature of their tasks. Failure of non-route nodes will not cause significant impact to the operation of IRIS, however if a route node stops functioning, IRIS must establish a new route from where the route breaks. As described earlier in sections III and IV, the previous hop of the failed node becomes a route-end node and starts to look for a node to join the route. To evaluate how IRIS recovers from route node failure, we simulate a random route node failure after a route is found and record the time that a new route is established. Fig. 23 shows the route recovery time of 494 node failures simulated with 90 randomly found routes of all three randomly constructed networks (30 routes per network). A clear relationship between the route recovery time and the distance of the failed node to the end base station can be observed. It takes longer time for the route to recover if the failed node is far away from the end base station, which is almost equivalent to rebuilding the route.

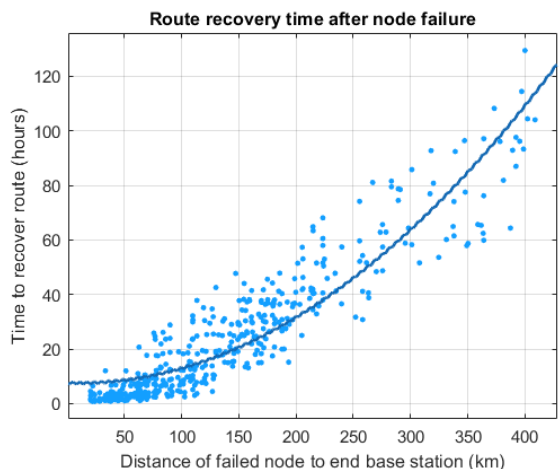

Fig. 23 Route recovery time from node failure

### F. Route Formation Time and Duty Cycle

The IRIS protocol is designed to operate with flexible duty cycle configurations which could be affected by both spectrum regulations and node energy budget. For example, the previous simulations targeting 1% duty cycle (active)have not exploited all energy budget (affordable duty cycle configurations are shown in Table III) provided by the energy harvesting device. Fig. 24 shows the route formation time of different slots-to-listen (STL) values per frame. Significant route formation time improvements can be observed when using larger STL (the number of hops are similar across different STL values). When exploiting more energy budget (e.g. when STL is set to 12) 90% simulations only require 15 hours to find routes, which is 270 frames. On the other hand, IRIS can support devices with lower energy budget by tuning down the duty cycle, resulting longer route formation time. The flexibility of duty cycle configurations allows the IRIS protocol to be applicable to networks with different energy budgets and spectrum regulations.

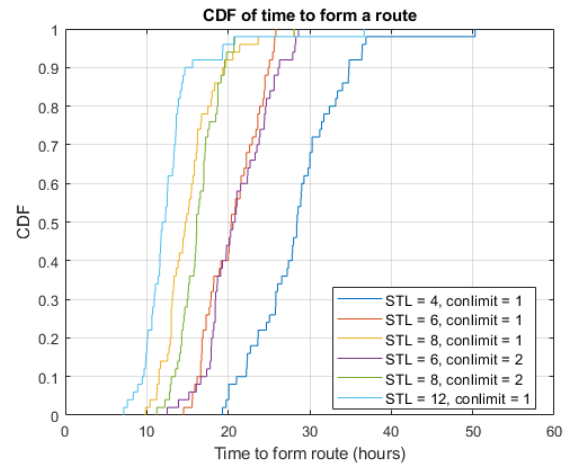

Fig. 24 Faster route formation time

## VI. CONCLUSION

This paper has presented IRIS, a unique low duty cycle cross-layer protocol designed for long-range WSNs with extremely limited power budget. Periodic ping packets are used to find the route to the base station as well as carry monitoring information. The main features of the IRIS protocol include its simplicity, the consistent low energy cost across all operation conditions and robustness against unreliable wireless connections and devices. The major contribution which makes the IRIS advances the state-of-the-art is that the IRIS does not require prior network information or node geolocation to provide route formation and data delivery with flexible duty cycle configurations. A large number of Monte Carlo simulations have been conducted to find the appropriate parameters to operate the IRIS protocol. Simulations results have shown guaranteed route formation under various conditions and robust route recovery after route node failures.

### ACKNOWLEDGMENT

The research was supported by TE Connectivity who has provided advice on applications and implementations. Monte Carlo simulations were conducted by the York Advanced Research Computing Cluster, which provided the platform of executing massive number of Matlab simulations in parallel.